\documentclass[conference]{IEEEtran}
\IEEEoverridecommandlockouts
\usepackage{cite}
\usepackage{amsmath,amssymb,amsfonts}
\usepackage{algorithmic}
\usepackage{graphicx}
\usepackage{textcomp}
\usepackage{xcolor}
\usepackage{makecell}
\def\BibTeX{{\rm B\kern-.05em{\sc i\kern-.025em b}\kern-.08em
    T\kern-.1667em\lower.7ex\hbox{E}\kern-.125emX}}

\usepackage{array,multirow}
\usepackage{booktabs}
\usepackage{color, colortbl}
\definecolor{Gray}{gray}{0.9}
\PassOptionsToPackage{bookmarks=false}{hyperref}

\begin{document}

\title{Self-supervised Attention Model for Weakly Labeled Audio Event Classification}

\author{\IEEEauthorblockN{Bongjun Kim}
\IEEEauthorblockA{
\textit{Northwestern University}, 
Evanston, IL, USA \\
bongjun@u.northwestern.edu}
\and
\IEEEauthorblockN{Shabnam Ghaffarzadegan}
\IEEEauthorblockA{
\textit{Bosch Research and Technology Center}, 
Sunnyvale, CA, USA\\
shabnam.ghaffarzadegan@us.bosch.com}
}

\maketitle

\begin{abstract}
We describe a novel weakly labeled Audio Event Classification approach based on a self-supervised attention model. The weakly labeled framework is used to eliminate the need for expensive data labeling procedure and self-supervised attention is deployed to help a model distinguish between relevant and irrelevant parts of a weakly labeled audio clip in a more effective manner compared to prior attention models. We also propose a highly effective strongly supervised attention model when strong labels are available. This model also serves as an upper bound for the self-supervised model. The performances of the model with self-supervised attention training are comparable to the strongly supervised one which is trained using strong labels. We show that our self-supervised attention method is especially beneficial for short audio events. 
We achieve 8.8\% and 17.6\% relative mean average precision improvements over the current state-of-the-art systems for \textit{SL-DCASE-17} and \textit{balanced AudioSet}.
\end{abstract}


\section{Introduction}
Audio event classification (AEC) with the goal of understanding the environment through sound recognition is one of the growing topics in research \cite{gemmeke2017audio,Takahashi2016,Zhuang,45611}. As deep learning approaches have been successfully applied to AEC, having access to a large amount of labeled data becomes very important. However, creating strongly labeled data with the exact time boundaries of audio events is a labor intensive task. Moreover, human annotation is error-prone due to the ambiguity in the beginning and end time and short duration of some audio events \cite{krizhevsky2012imagenet}.

One way to relax the need for strong labels is to address AEC problem in a weakly labeled framework in which time information of an event is not required. A common way to train a model on weakly labeled data is to feed a set of audio clips as the model input and minimize clip-level loss computed by pooling operations over time segments \cite{kumar2017audio}. This approach has been used in many deep learning-based solutions \cite{kumar2017deep,DBLP:conf/icassp/WangM17,parascandolo2017convolutional,DBLP:journals/corr/abs-1804-10070}. Recently, attention scheme has been applied to weakly labeled AEC. The attention mechanism helps a model focus on sub-sections of audio which contribute to the classification while ignoring the irrelevant instances such as background noises \cite{Chen:2018:CSA:3206025.3206067,ijcai2018-463}.

The attention mechanism is usually implemented by adding extra layers into a neural network model. The model is then trained by minimizing loss between clip-level outputs pooled through the attention mechanism and weak labels. While these models have proved to be effective for weakly labeled AEC \cite{DBLP:journals/corr/abs-1711-00927,Chen:2018:CSA:3206025.3206067}, we propose a novel solution to train the attention layers in a more effective way to better tune the attention matrix. The idea is to add direct supervision to the attention layer. The supervision is performed by strong labels (i.e. strong supervision) as well as the network itself only with weak labels (i.e. self-supervision) during training. Our model allows simple integration of these two supervisions in one framework.

As far as we know, our paper is the first work using direct supervision to attention learning for weakly labeled AEC while there is a recent concurrent work \cite{li2018tell} using the self-supervision but for image segmentation. The difference is that while their model takes a fixed size of images, our attention model can deal with a variable length of audio, which is very important for AEC. Moreover, attention for audio recognition might be more challenging than attention for image recognition in some cases because different classes of sound events can be fully overlapped, but they still need to be identified. There is also another prior work \cite{ijcai2018-463} which named the proposed method attentional supervision. But they use attention scores to obtain segment-level predictions and compute segment-level loss with weak labels. They do not have any direct supervision to better tune the attention matrix.

Our main contributions are followings: 1) we propose an effective strongly-supervised attention mechanism method for strongly labeled AEC task; 2) we propose self-supervised attention mechanism as a new training scheme for weakly labeled AEC task; 3) we integrate strong and self-supervised attention schemes in one general framework to enable using weak and strong labels (if available) at the same time. We show the proposed models outperform existing architectures for AEC problems for both strongly and weakly labeled data.

\section{Strongly supervised attention model}
\label{sec:Strongly_supervised}

\begin{figure*}[t]
  \centering
  \includegraphics[width=\textwidth, height=3.15cm,,keepaspectratio]{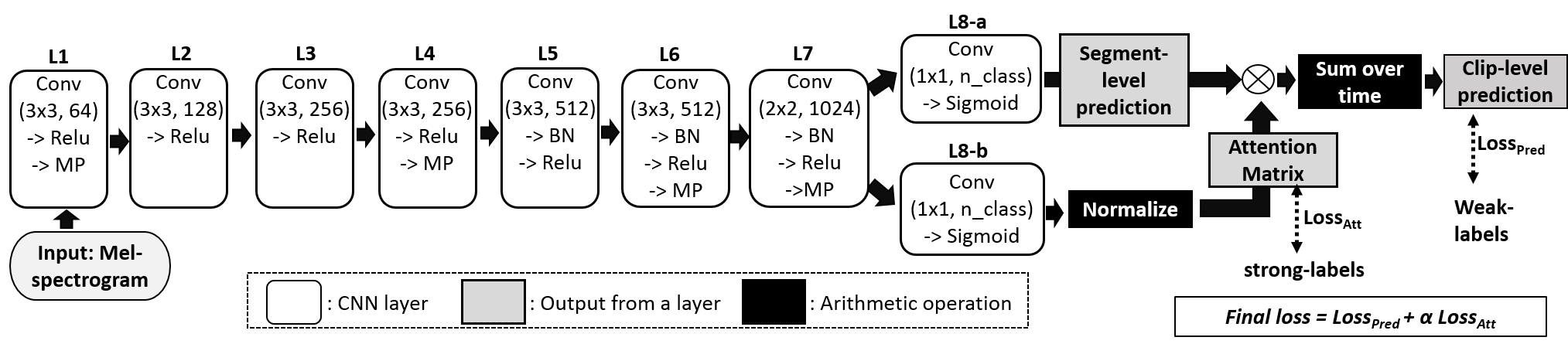}
  \caption{Proposed network architecture with strongly supervised attention model.}
  \label{fig:network_architecture}
\end{figure*}

In this section, we present a strongly supervised attention model with two purposes: 1) exploring best possible ways of incorporating strong labels in training; 2) as an upper-bound performance for our proposed self-supervised attention architecture. The two models share the same architecture components while having different loss functions.
Figure~\ref{fig:network_architecture} demonstrates the proposed model architecture inspired by \cite{45611,kumar2018knowledge,DBLP:journals/corr/abs-1711-00927}. Our model is a fully convolutional network. Details of each layer configuration are explained in section \ref{sec:experiments}. 

Clip-level predictions with attention are made by two layers of L8, L8-a and L8-b. Suppose the outputs from L8-a and L8-b are $X^{a}, X^{b}\in\mathbb{R}^{C\times T}$ where $C$ is the number of classes and $T$ is the number of time segments. The number of segments $T$ depends on the input audio length. $X^{a}$ contains class probabilities for each time segment (\textit{Segment-level prediction} in Figure \ref{fig:network_architecture}). Layer L8-b is an attention layer that learns which cells in the segment-level prediction ($X^{a}$) are the most relevant ones when the segments are pooled over time. We call a matrix containing class-wise attention weights for each time segment, \textit{attention matrix} $A\in\mathbb{R}^{C\times T}$. This matrix is computed by normalizing the output from L8-b ($X^{b}$) as $A = X^{b} / \sum_{i=1}^{T} X^{b}_{i}$, where $X^{b}_{i}\in\mathbb{R}^{C}$ is $i$th temporal segment of $X^{b}$.
Once the attention matrix $A$ is computed, clip-level prediction is performed by: $\hat{y} = \sum_{i=1}^{T}A_{i} \odot X^{a}_{i}$, where $\hat{y}\in\mathbb{R}^{C}$ is a clip-level prediction, $X^{a}_i$ is $i$th temporal segment of $X^{a}$, and the notation $\odot$ indicates element-wise multiplication.

\textbf{Strongly supervised attention loss}: Existing attention models for weakly labeled AEC are usually trained by minimizing loss between clip-level labels, $y\in\{0, 1\}^{C}$, and the clip-level predictions \cite{DBLP:journals/corr/abs-1711-00927,ijcai2018-463}. The attention matrix learned in this process will focus on the most \textit{relevant} and \textit{discriminative} parts of the audio clip for prediction. Our model is trained with direct supervision to attention matrix, minimizing the sum of two loss terms, prediction loss $L_{pred}$ and attention loss $L_{att}$. Attention model trained with strong supervision has two advantages: 1) Strong supervision helps the attention matrix to cover all the \textit{relevant} parts of the clip rather than the most \textit{relevant} and \textit{unique} parts as it is in other existing attention models. We believe focusing on larger regions with possible contribution will enhance the final system prediction. 
2) {\color{black} It allows a model to utilize both weakly and strongly labeled data simultaneously within one framework.}

The prediction loss is binary cross-entropy loss between weak labels and clip-level predictions as, $L_{pred} = -\sum_{c=1}^{C} \hat{y}^{c} log y^{c} + (1 - \hat{y}^{c}) log (1-y^{c})$,
where $\hat{y}^{c}$ is the likelihood of event class $c$ and $y^{c}$ is ground truth label for the clip. To compute attention loss $L_{att}$, segment-level labels are needed to guide the attention matrix $A$. If the strong labels (time boundary information) are available, one can simply obtain the segment-level labels by calculating the size of receptive fields for each segment. Suppose $Y\in\{0, 1\}^{C \times T}$ is a matrix containing ground truth labels for each time segment. Then, $L_{att}$ is computed as, $L_{att} = -\sum_{i=1}^{T}\sum_{c=1}^{C} A_{i}^{c} log Y_{i}^{c} + (1 -A_{i}^{c}) log (1-Y_{i}^{c})$,
where $A_{i}^{c}$ is an attention weight of event class $c$ for $i$th segment and $Y_{i}^{c}$ is a ground truth label of event class $c$ for $i$th segment of the clip. Then our final loss can be obtained as $L = L_{pred} + \alpha L_{att}$, where $\alpha$ is a hyper-parameter that weights the supervision to attention matrix. 

\begin{figure}[b]
  \centering
  \includegraphics[scale=0.42]{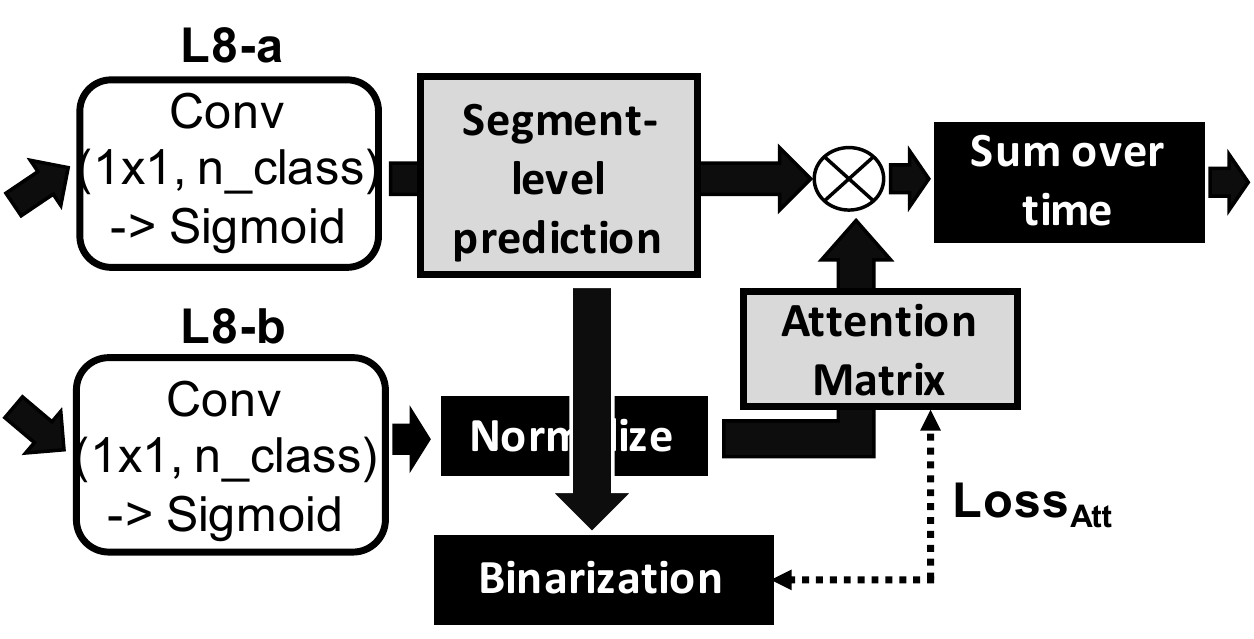}
  \caption{Self-supervised attention model. During training, segment-level prediction is binarized on-the-fly and attention loss is computed between the \textit{psuedo}-strong label and attention matrix.}
  \label{fig:online_pseudo_gen}
\end{figure}

\section{Self-supervised attention model}
In this section, we propose a way of leveraging the benefits of the strongly supervised attention model when only weak labels are available. The idea is to generate pseudo-strong labels using the trained models in every forward pass of training. While the self-supervision by the pseudo-strong labels might not be accurate during early rounds of training, they become more accurate as the model converges over many iterations.

Figure \ref{fig:online_pseudo_gen} shows how the self-supervision works during training. To generate pseudo-strong labels to be used for the self-supervision during training, segment-level prediction (i.e., output from L8-a) is computed and binarized in every forward passes of training. The binarization is performed as
\begin{equation}\label{eq:binarization}
P_{c,t}=
\begin{cases}
0, & \text{if}\ (c\not\in C) \vee (X_{c, t}^{a} < \theta)\\
1, & \text{otherwise,}
\end{cases}
\end{equation}
where $P_{c, t}$ is a pseudo-strong label for $c$-th event at time $t$, $X_{c, t}^{a}$ is output from L8-a (segment-level prediction) which is a likelihood of $c$-th event at time $t$, C is a set of entire classes, and $\theta$ is a threshold for binarization. We set the threshold by averaging the elements in $X^{a}$ that are \textit{irrelevant} to classes presented in the audio clip. Once the pseudo-strong labels are generated, self-supervised attention loss is computed in the same way the strongly supervised attention loss is computed, but with the pseudo-strong labels. Note that self-supervision attention loss can be easily added as a third term to the strongly supervised attention loss in Figure \ref{fig:network_architecture} to combine the benefits of both models or as model adaptation when a few strong labels from a new domain are available.

\section{Experiments}
\label{sec:experiments}
We performed experiments to confirm two hypothesises: 1) the strongly supervised attention model is the most effective model for audio classification tasks when strong labels are available during training, 2) the self-supervised attention model is an effective way for audio classification tasks when only weakly labels are available. 

\begin{figure}[t]
  \centering
  \includegraphics[scale=0.42]{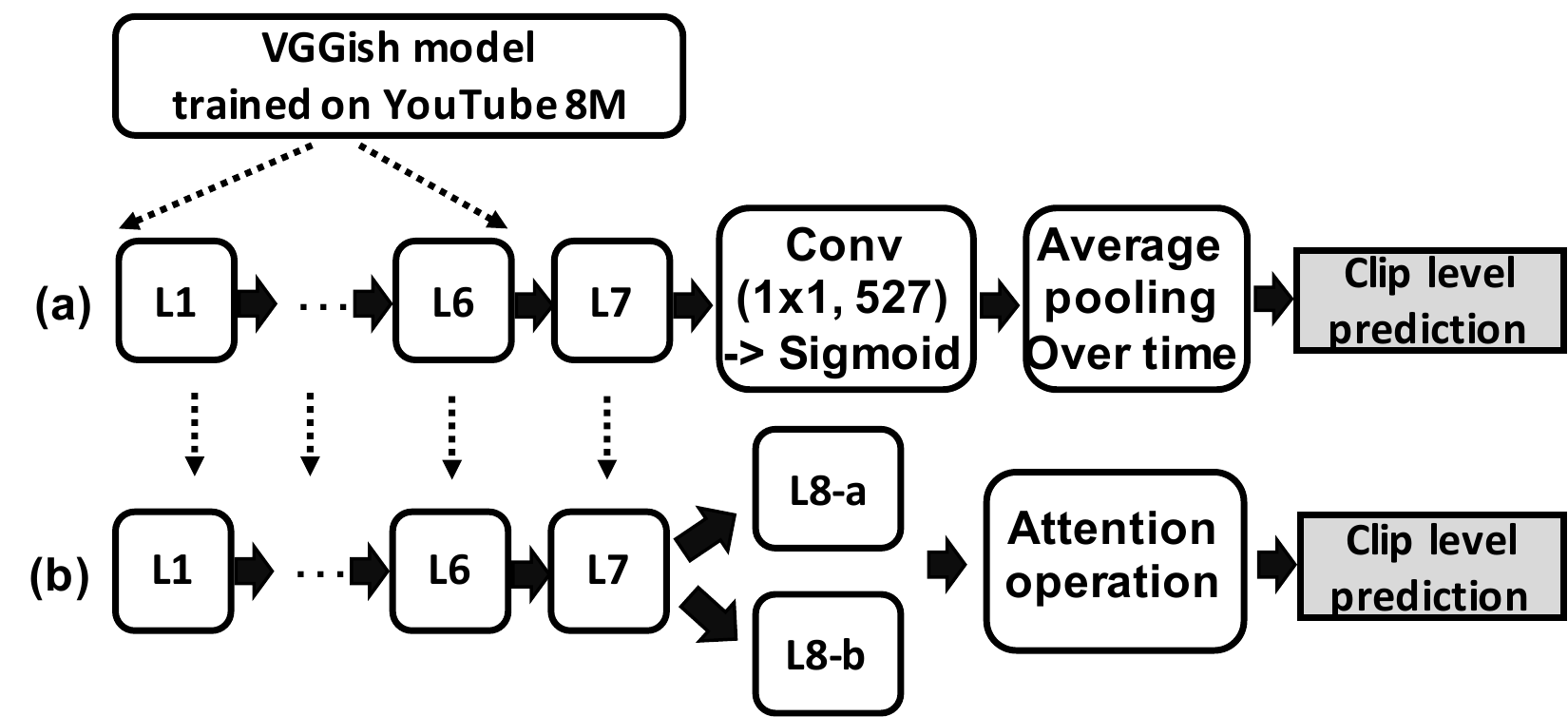}
  \caption{Two steps of transfer learning: 1) Pre-trained parameters from \textit{VGGish} model trained on \textit{YouTube 8M} data are transferred to layers L1 to L6 of model(a). Model(a) is fine-tuned on \textit{AudioSet} with 527 classes. 2) The tuned parameters from L1 to L7 of Model(a) are transferred to model(b) and fine-tuned on \textit{SL-DCASE-17} with 17 classes. Model(b) is equivalent to the model in Figure \ref{fig:network_architecture}.}
  \label{fig:transfer}
\end{figure}


\textbf{Dataset and performance metric}: We evaluate our models on two subsets of Google AudioSet \cite{gemmeke2017audio}: \textit{SL-DCASE-17} and \textit{balanced AudioSet}. 
AudioSet is a large scale weakly labeled data collected from YouTube videos consisting of 2 millions of 10-second audio clips with 527 classes. The dataset is pre-divided into three subsets of balanced training (Bal-train), unbalanced training (Unbal-train) and evaluation (Eval) sets.

\textit{SL-DCASE-17}: Evaluating the strongly supervised and self-supervised attention mechanisms requires both strong and weak labeled datasets. Therefore, We use evaluation and test sets of DCASE 2017 challenge (Task 4) \cite{mesaros:hal-01627981} where strong labels are available. This dataset is a subset of Google AudioSet with 17 classes. As a result, we created our own strongly-labeled training set by combining the evaluation and test sets of this challenge, with a total of 1,591 recordings. For validation set, we sample 1,638 recordings ($\sim$100 recordings per class) from Unbal-train of AudioSet. Our test set is sampled from Eval of AudioSet with 626 audio recordings of the 17 classes. We call our new partitioned dataset \textit{SL-DCASE-17} in which SL stands for Strongly Labeled. We have released the list of YouTube IDs used in \textit{SL-DCASE-17} \footnote{https://github.com/bongjun/eusipco19} for future references.

\textit{Balanced AudioSet}:
Given that \textit{SL-DCASE-17} dataset does not follow the standard challenge partitioning for train/test/validation sets, we use \textit{Balanced AudioSet} for a fair comparison to prior works. However, we are only able to evaluate our self-supervised model due to the lack of strong labels in this data. We train and evaluate our model on the Bal-train and Eval sets of AudioSet including entire 527 classes. 10\% of the Bal-train is used for validation. Finally, we have 18,677 of training set, 2,262 of validation set, and 19,330 of testing set.
We choose balanced AudioSet instead of unbalanced one due to our resource limitations. Also note that validation sets in all our experiments for both \textit{SL-DCASE-17} and \textit{Balanced AudioSet} datasets are used for adjusting the learning rate, $\alpha$ and choosing the best model. No other parameter tuning such as number of layers and/or nodes are done. For performance metric, we compute mean Average Precision (mAP) \cite{Buckley:2004:REI:1008992.1009000} over all classes, as it is used in most of the prior works and DCASE 2017-2018 challenges.

\textbf{Training settings}:
All audio files are resampled to 16kHz mono. Each sample is represented by a log-scale Mel-spectrogram with 64 Mel bins, a window size of 25 ms and hop size of 10 ms. Although the proposed models take an arbitrary length of an input representation, we fix the input dimensions to $998 \times 64$ ($\sim$10 seconds of audio) for batch training, by zero-padding or truncating the signal. During inference, variable signal lengths are fed to the network.

The network configuration is shown in Figure \ref{fig:network_architecture}. Note that the number of filters on L8 depends on the number of classes for each task. To find the best $\alpha$ value (weight for attention loss) in the loss function presented in section \ref{sec:Strongly_supervised}, we performed a grid-search in the range of 0.1 to 1.5 with 0.1 step size. Each experiment was repeated 5 times, with $\alpha=1$ having the best performance on average. Adam Optimizer with learning rates of 0.001 for \textit{AudioSet} and 0.0001 for \textit{SL-DCASE-17} are used for training. The best models on the validation set during the first 50 epochs are chosen for testing. We train each model 20 times to compare variants of models statistically.

\textbf{Transfer learning}
While training deep neural networks usually requires a large amount of training data, our generated \textit{SL-DCASE-17} data contains only 1,591 training samples. To overcome this issue, we apply transfer learning to our models in two steps. First, we configure a model with the same architecture as the proposed model except for the last attention layer. Figure \ref{fig:transfer}-(a) shows the architecture where attention layers are replaced with an average pooling layer to output the clip-level prediction. Then, layers L1 to L6 of the model(a) are initialized with the \textit{VGGish} model \cite{45611} which has been pre-trained on the \textit{YouTube 8M} dataset \cite{abu2016youtube}. Next, the model(a) is fine-tuned on Bal-train of \textit{AudioSet} which has only weak labels. This step is done to adapt a YouTube 8M-\textit{VGGish} model, a video related task, to AudioSet model, an audio related task. Next, we select the best model on the validation set of Bal-train and transfer the trained weights to our attention model, Model(b) in Figure \ref{fig:transfer}. It is then fine-tuned on \textit{SL-DCASE-17} for better representation of the specific 17 classes defined in this dataset. Note that the testing examples for the final target tasks on \textit{SL-DCASE-17} are not used during any pre-training steps. We call this model ``Two-step transfer'' model. We also examine ``One-step transfer'' model where knowledge is transferred directly from \textit{VGGish} model to our attention model without adapting on a larger set of Bal-train of \textit{Audioset}.

\begin{table}[t]
  \caption{Classification performance on \textit{SL-DCASE-17} test set from 20 trials with one/two-step transfer learning.}
  \label{fig:transfer1}
  \centering
  \resizebox{\columnwidth}{!}{%
  \begin{tabular}{ccccc||c|c} 
    \toprule
    & \textbf{Attention} & & & & \multicolumn{1}{c}{\textbf{One-step Trans}.} & \multicolumn{1}{c}{\textbf{Two-step Trans}.} \\ \cline{1-3}
    \makecell{No \\Supervision} & \makecell{Strongly \\ supervised} & \makecell{Self\_\\supervised} & \makecell{Strong\\label} & \makecell{Weak \\label} & mAP &  mAP \\
    \hline
     &  &  &  & \checkmark & 50.0$\pm$0.33 & 51.4$\pm$0.17 \\
     \hline
     &  &  & \checkmark &  & 49.7$\pm$0.42 & 50.8$\pm$0.44\\
     \hline
     &  &  & \checkmark & \checkmark & 49.0$\pm$0.44 & 51.1$\pm$0.31\\
     \hline
    \checkmark &  &  &  & \checkmark & 53.8$\pm$0.44 & 53.9$\pm$0.47\\
    \hline
    \rowcolor{Gray}
     & \checkmark &  & \checkmark &  & \textbf{54.7$\pm$0.4} & \textbf{56.3$\pm$0.26}\\
     \hline
     \rowcolor{Gray}
     &  & \checkmark &  & \checkmark & \textbf{54.4$\pm$0.12} & \textbf{55.9$\pm$0.30}\\
	\bottomrule
     \end{tabular}
     }
\end{table}

\begin{figure}[!b]
  \centering
  \includegraphics[scale=0.26]{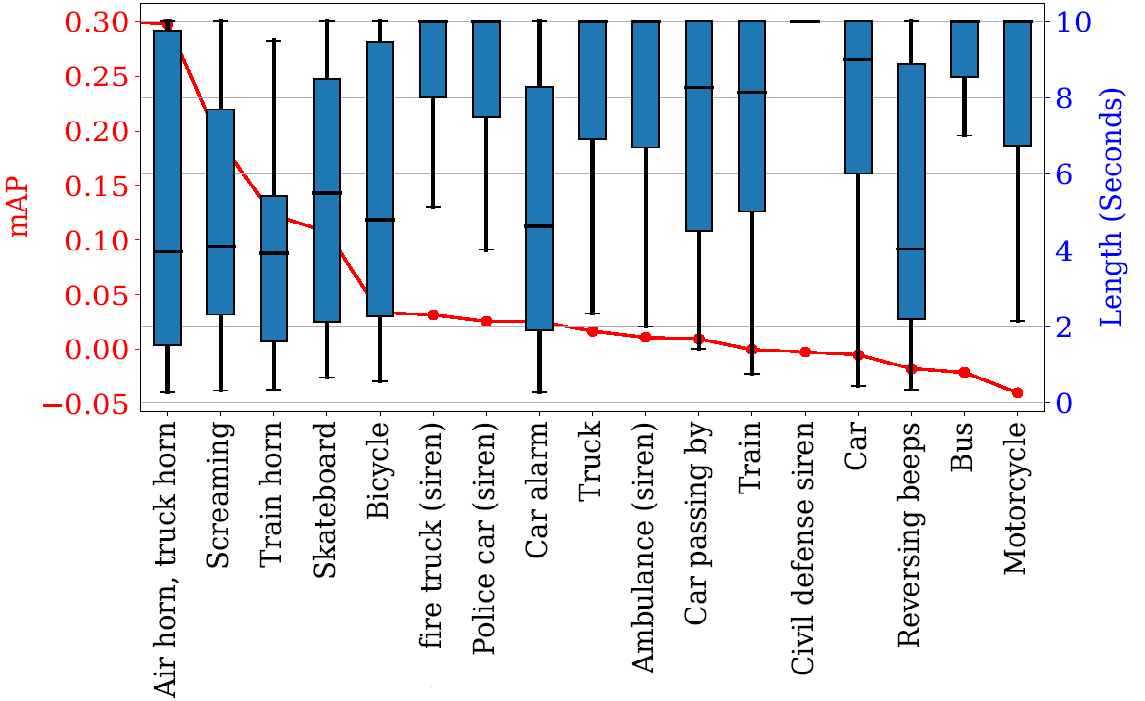}
  \caption{Class-wise absolute mAP improvement of unsupervised and self-supervised attention models (4th and 6th rows of Table \ref{fig:transfer1}) vs. time duration of each class in seconds.}
  \label{fig:plot3}
\end{figure}

\textbf{Results-SL-DCASE-17:}
To evaluate the effectiveness of our proposed solution on \textit{SL-DCASE-17}, We compare six different models with combinations of one-step/two-step transfer, strong/weak labels, non-supervised/strongly-supervised/self-supervised attention models. Table \ref{fig:transfer1} presents mAPs of 20 trials of each variant of the models. Results from our models are highlighted in gray. Note that our task is detecting the presence of an audio event within a clip (i.e. audio tagging). The model with no attention and with strong labels (2nd row of Table \ref{fig:transfer1}) is trained by minimizing loss between segment-level predictions and strong labels without pooling operation. And the model with no attention and with weak and strong labels (3rd row of the tables) is trained by minimizing the sum of segment-level loss before pooling and clip-level loss after average pooling. Even though both models are trained using strong labels, they still rely on pooling operation to output the clip-level prediction during inference.

We can firstly confirm, in Table \ref{fig:transfer1} the merit of using two-step transfer model. A statistical test (Wilcoxon signed-rank test) with 20 trials shows that the performance gain by two-step transfer learning is statistically significant ($p<0.05$) for all the models except for the model with unsupervised attention and weak labels (4th row of the tables). Table \ref{fig:transfer1} also shows that attention models outperform non-attention models. Specifically, we observe with no attention mechanism, strong labels have no benefit for the audio tagging task. As expected, attention models trained on only weak labels without any supervision outperforms models trained with strong labels by $\sim$3\% absolute mAP. As discussed in \cite{DBLP:journals/corr/abs-1804-09288}, incorporating attention mechanism in weak labeling scenarios especially with low density labels is crucial which is often the case in audio applications. The density of labels here means the portion of the recording in which the audio event exists out of the total length. Most importantly, we observe that both our attention models with strong and self-supervision increase the system performance considerably (Wilcoxon signed-rank test, $p<0.05$). Self-supervised model with no strong labels has almost on-par performance with the strongly supervised attention. This is a valuable result for us since in most of the audio applications only weak labels are available. 

\begin{figure}[!b]
  \centering
  \includegraphics[width=0.86\linewidth]{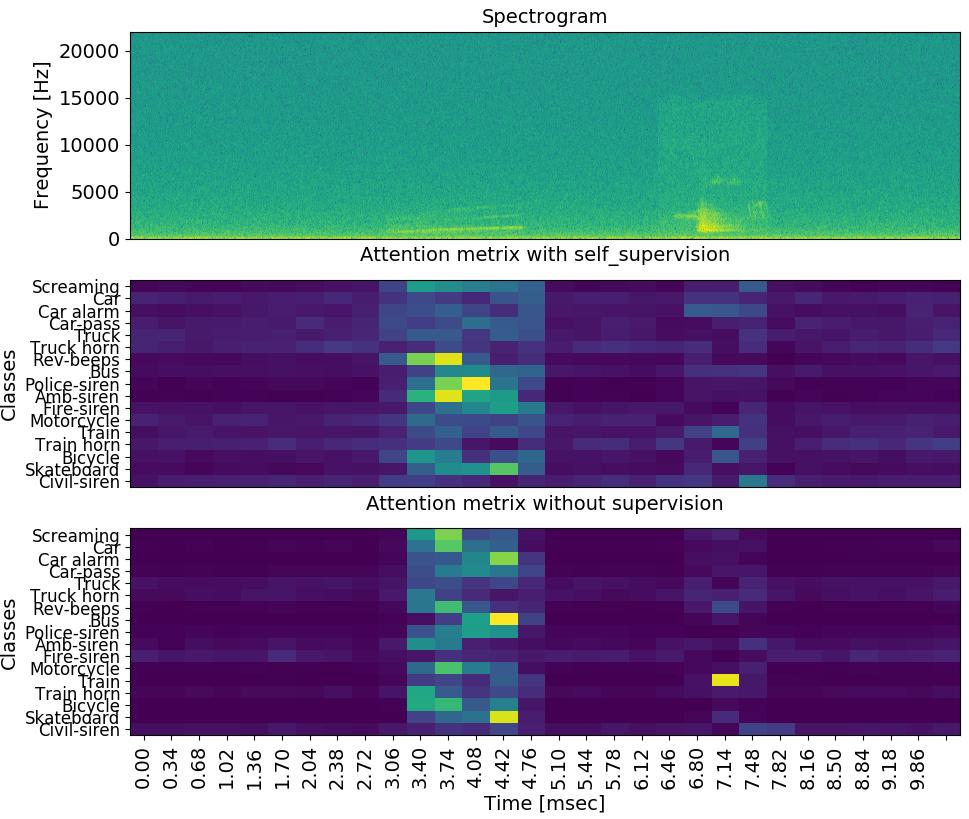}
  \caption{Attention matrix for an audio file with siren (around 4 sec) and dog bark (around 7 sec). The self-supervised attention matrix covers wider areas and more relevant classes. It has smoother transitions in the start and end of audio events. Color bar of attention matrices is from blue to yellow with dark blue having the lowest value and dark yellow having the highest value.}
  \label{fig:attn_siren}
\end{figure}

For comparison, we have also calculated the performance of the winning solution of DCASE17 Task4 challenge \cite{Xu2017}. We have tested their released model on our test set. Their best model reaches 53\% mAP. Note that some of our testing examples are included in the training set of task4. As a result, the best model of DCASE17 has been exposed to some of our testing examples during its training. However, our solution still outperforms it with 3\% mAP absolute improvement.

To examine which classes benefit the most from supervision on attention matrix, detailed class-wise absolute mAP improvements are shown in Figure \ref{fig:plot3} between unsupervised and self-supervised attention models (4th and 6th rows of Table \ref{fig:transfer1}) for \textit{SL-DCASE-17}. We see that the self-supervised attention model outperforms the unsupervised one for the majority of the 17 classes. Especially, for shorter ones such as ``Air horn, truck horn", ``screaming" and ``train horn". This experiment confirms one more time the importance of an accurate attention mechanism especially for low density weak labels. 

Figure \ref{fig:attn_siren} shows a spectrogram of an audio clip (first row) with siren and dog bark events, along with its attention matrices during inference. As shown in the figure, self-supervised attention matrix has better estimates of time boundaries as well as relevant classes, with events such as police siren, ambulance sirens and reversing beeps with high values around the ground truth siren segment. Moreover, time boundaries are smoother for the detected events (transitional colors in the start and end of the event). Also, given that dog bark is not one of the target classes in \textit{SL-DCASE-17}, the self-supervised model assigns very low values to its corresponding time segment while the unsupervised model has trouble filtering out this event.



\begin{table}[!t]
  \caption{Sound event classification on balanced AudioSet. Our results are from 20 trials.}
  \label{fig:audio-bal}
  \centering    
  \resizebox{\columnwidth}{!}{%
  \begin{tabular}{cccc||cc} 
    \toprule
    & \multicolumn{2}{c}{\textbf{ Attention}.} & & &  \\ \cline{2-3}
    Methods & \makecell{No \\Supervision} & \makecell{Self\_\\supervised} & \makecell{Weak \\label} & mAP & \#Params\\
    \hline
    TLWeak \cite{kumar2018knowledge} & & & \checkmark & 21.3 & 5M \\
    \hline
    ResNet-ATT \cite{xu2017attention,ijcai2018-463} & \checkmark &   &  \checkmark & 22.0 & 4M\\
    \hline
    M\&mnet-MS \cite{ijcai2018-463}& \checkmark & & \checkmark & 23.3 & 8M \\
    \bottomrule
    No attention &  &  &  \checkmark & 26.4$\pm$0.01 & 6.7M\\
    \hline
    Unsupervised attention & \checkmark &  & \checkmark & 23.1$\pm$0.06 & 7.3M \\
    \hline
    \rowcolor{Gray}
    Self-supervised attention &  & \checkmark & \checkmark & \textbf{27.4$\pm$0.02} & 7.3M\\
	\bottomrule
     \end{tabular}
     }
\end{table}

\textbf{Results-Balanced AudioSet:}
We evaluate the models on \textit{Balanced AudioSet} to validate our proposed solution on larger scale data with a larger number of classes and also compare the results to the ones reported in prior works. Table \ref{fig:audio-bal} lists mAP and the number of trainable parameters of three existing models in the prior works, and 3 versions of our proposed models: 1) no attention, 2) unsupervised attention, 3) self-supervised attention. All these models only use weak labels for training since strong labels are not available for this dataset. We report the mean performance of 20 trials for our models. Note that all the prior works reported in Table \ref{fig:audio-bal} also use the same train/test/validations sets as our work (Balanced train of \textit{Audioset}). However, they did not perform any statistical test while we did using the 20 trials.

As shown in the table, our first model without attention outperforms all existing models. This performance gain comes from the effect of transfer learning by \textit{VGGish} model. Through experiments, we found that fixing the first three layers of VGGish and fine-tuning other layers work best. An interesting point in these results is that adding attention mechanism without supervision hurts the system performance. We believe attention matrix without supervision fails to learn the relevant regions due to very low-density labels of many classes. Our self-supervised model achieved the best performance on \textit{Balanced AudioSet}. The statistical test (Wilcoxon signed-rank test) confirms that it significantly outperforms both non-attention and unsupervised attention models ($p<0.05$).


\section{Conclusion}
\label{sec:Conclusion}
We present a new framework to train an attention module in strongly supervised and self-supervised manners for an AEC task when a few or no strong labeled data are available. In the strongly supervised attention model, the attention layer is directly updated via strong labels, while the self-supervised attention model uses pseudo-strong labels which are generated only by weak labels during training. The experiments confirm that the direct supervision on attention matrix with strong labels and/or weak labels is a very effective way of training attention models for audio even classification.


\bibliographystyle{IEEEbib}
{\footnotesize
\bibliography{refs}
 }

\end{document}